\documentclass[aps,pra,preprint,amsmath,amssymb]{revtex4}
\usepackage{epsfig,epsf}
\usepackage{dcolumn}
\usepackage{bm}
\usepackage{graphics,psfrag}
\usepackage{graphicx,psfrag}

\begin{document}
\newcommand{\ri}{{\rm i}}
\newcommand{\re}{{\rm e}}
\newcommand{\bx}{{\bf x}}
\newcommand{\bb}{{\bf b}}
\newcommand{\bd}{{\bf d}}
\newcommand{\be}{{\bf e}}
\newcommand{\br}{{\bf r}}
\newcommand{\bk}{{\bf k}}
\newcommand{\bB}{{\bf B}}
\newcommand{\bE}{{\bf E}}
\newcommand{\bI}{{\bf I}}
\newcommand{\bJ}{{\bf J}}
\newcommand{\bR}{{\bf R}}
\newcommand{\bS}{{\bf S}}
\newcommand{\cL}{{\cal L}}
\def\Jp#1{J_+^{(#1)}}
\def\Jm#1{J_-^{(#1)}}
\newcommand{\bZero}{{\bf 0}}
\newcommand{\bM}{{\bf M}}
\newcommand{\bn}{{\bf n}}
\newcommand{\bs}{{\bf s}}
\newcommand{\tbs}{\tilde{\bf s}}
\newcommand{\rSi}{{\rm Si}}
\newcommand{\beps}{\mbox{\boldmath{$\epsilon$}}}
\newcommand{\bmu}{\mbox{\boldmath{$\mu$}}}
\newcommand{\rg}{{\rm g}}
\newcommand{\tr}{{\rm tr}}
\newcommand{\xmax}{x_{\rm max}}
\newcommand{\ra}{{\rm a}}
\newcommand{\rx}{{\rm x}}
\newcommand{\rs}{{\rm s}}
\newcommand{\rP}{{\rm P}}
\newcommand{\up}{\uparrow}
\newcommand{\down}{\downarrow}
\newcommand{\hc}{H_{\rm cond}}
\newcommand{\kb}{k_{\rm B}}
\newcommand{\cI}{{\cal I}}
\newcommand{\tit}{\tilde{t}}
\newcommand{\cE}{{\cal E}}
\newcommand{\dgtwo}{\langle \delta g^2\rangle}
\newcommand{\dgthree}{\langle \delta g^3\rangle}
\newcommand{\cC}{{\cal C}}
\newcommand{\Ubs}{U_{\rm BS}}
\newcommand{\qq}{{\bf ???}}
\newcommand*{\etal}{\textit{et al.}}
\def\vec#1{\mathbf{#1}}
\def\ket#1{|#1\rangle}
\def\bra#1{\langle#1|}
\def\keps{\mathbf{k}\boldsymbol{\varepsilon}}
\def\dm{\boldsymbol{\wp}}
\def\CG#1#2#3#4#5#6{C{\small \begin{array}{ccc}{#1}&{#3}&{#5}\\{#2}&{#4}&{#6}\end{array}}}
\def\cLL#1#2#3#4#5#6#7#8{L{\small \begin{array}{cccc}{#1}&{#3}&{#5}&{#7}\\{#2}&{#4}&{#6}&#8\end{array}}}
\def\CGprim#1#2#3#4#5#6{C^{{#1}\,{#3}\,{#5}}_{{#2}\,{#4}\,{#6}}}
\def\CLLprim#1#2#3#4#5#6#7#8{{\cal L}_1^{{#1}\,{#3}\,{#5}\,{#7}}_{{#2}\,{#4}\,{#6}\,{#8}}}

\sloppy

\title{Superradiance of cold atoms coupled to a superconducting
  circuit}
\author{Daniel Braun $^{(1,2)}$, Jonathan Hoffman $^{(3)}$, and Eite
  Tiesinga $^{(3)}$}
\affiliation{{$^{(1)}$ Laboratoire de Physique Th\'eorique  -- IRSAMC,
  Universit\'e de Toulouse, UPS, F-31062 Toulouse, France,}\\ 
$^{(2)}$ LPT -- IRSAMC, CNRS, F-31062 Toulouse, France,}
\affiliation{$^{(3)}$ Joint Quantum Institute,
National Institute of Standards and Technology and University of
Maryland,
100 Bureau Drive, Stop 8423 Gaithersburg, Maryland 20899-8423, USA}

\centerline{\today}
\begin{abstract}
We investigate superradiance of an ensemble of atoms coupled to an
integrated superconducting LC-circuit.  Particular attention is paid to
the effect of inhomogeneous coupling constants.  Combining perturbation
theory in the inhomogeneity and numerical simulations we show that
inhomogeneous coupling constants can significantly affect the superradiant
relaxation process. Incomplete relaxation terminating in ``dark states'' can
occur, from which the only escape is through individual spontaneous emission
on a much longer time scale. The relaxation dynamics can be significantly
accelerated or retarded, depending on the distribution of the coupling
constants. On the technical side, we also generalize the previously known
propagator of superradiance for identical
couplings in the completely symmetric sector to the full exponentially
large Hilbert space.
\end{abstract}
\maketitle

\section{Introduction}
With the advance of experimental quantum information processing, the need to
study ``hybrid quantum processors'' has arisen \cite{duan_long-distance_2001,petrosyan_quantum_2008,petrosyan_reversible_2009,rabl_hybrid_2006,srensen_capacitive_2004,tian_interfacing_2004,verdu_strong_2009,wallquist_hybrid_2009}.  In such a processor
different physical systems are used in order to exploit their respective
advantages, such as long coherence times versus fast processing times, or
fast propagation in the case of quantum communication.
A  natural candidate for a hybrid quantum processor is a system of
atoms  
coupled to superconducting circuits
\cite{petrosyan_reversible_2009}. Circuit-QED schemes 
are rapidly 
emerging as a promising new avenue towards scalable quantum computation.
These schemes combine the strong coupling and precise control of cavity-QED
systems  
with the scalability of integrated solid state circuits. Nevertheless, the
coherence times achieved so far are of the order of a $\mu$s for
transmon qubits \cite{Houck08}. Much 
longer coherence times 
are achievable for superpositions of the hyperfine levels of alkali-metal
atoms, a fact well known from the design of atomic clocks.  Therefore, atoms are
predestined as a building block for a  
quantum memory.  The  coupling of the hyperfine levels to their environment
is through a magnetic dipole transition and thus much weaker than the
coupling through an electric dipole. This
limits the bandwidth with which information can be transferred to and from
such a quantum memory. It is beneficial to use a large number
of atoms to store a single excitation
\cite{duan_long-distance_2001}, as the 
single photon Rabi frequency that determines the rate of energy exchange
scales as $\sqrt{N}$ with the number of atoms $N$.  

Beyond their technical relevance, hybrid quantum processors are also
interesting for studying physical effects on a more fundamental level.
They allow for easily modifiable parameters, and to reach new parameter
regimes, so far inaccessible in traditional quantum optical systems. 
For
example, it has been proposed that the ultra-strong couplings of circuit-QED
schemes might allow one to observe the phase transition in superradiance
predicted last century by Mallory, and Hepp and Lieb
\cite{hepp_superradiant_1973,mallory_superradiant_1975,chen_simulation_2007}.
The existence of such a 
transition was subject of considerable theoretical debate (see
e.g.~\cite{rzazdotewski_phase_1975}), culminating in the recent claim that
the transition can in 
principle not be observed in cavity-QED, but should be observable in
circuit-QED
\cite{nataf_no-go_2010}.     
For atoms coupled to a superconducting
circuit one might envisage to perform precisely controlled experiments on
superradiance. Superradiance is a rather complex effect that can
include things like mode competitions or beating \cite{Gross82}. It would
therefore be desirable to go 
beyond previous experiments
\cite{Skribanowitz73,Gross76,Gross78,Gross79,Gross82} using 
``traditional'' cavity QED in terms of control of different parameters.

In this paper we study superradiance of an ensemble of atoms coupled to an
integrated resonant LC circuit.  We show that there is indeed a
regime in which superradiant behavior is expected.  At the same time,
additional complications arise due to variations in the coupling
constants, which are fundamentally related to the small size of the
integrated LC circuit. Most of this paper is therefore dedicated to the
study of the effects of inhomogeneous coupling constants on
superradiance. So far the study of superradiance has been almost
entirely limited to homogeneous coupling constants, and/or initial states
that are fully symmetric under all permutations of atoms.  
We develop a general theoretical framework that allows one to deal with such
inhomogeneities. Based on perturbation theory in the inhomogeneity, this
framework starts with the construction of a semiclassical solution of the
superradiant 
master equation  in the entire exponentially large Hilbert space for the
homogeneous case.

\section{Model}
\subsection{Physical system}\label{sec.physsys}
\begin{figure}
\epsfig{file=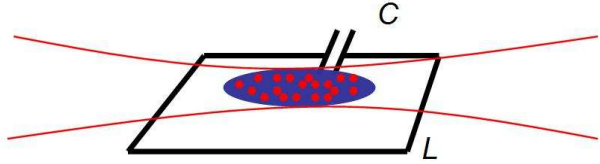,width=8cm,angle=0}
\caption{(Color online) $N$ two level atoms in a dipole trap with a
  transition in resonance 
  with a superconducting $LC$-circuit.
}\label{fig.exp}        
\end{figure}

We consider $N$ atoms that are, on the time scale of the experiment, held at
fixed positions close to an $LC$-circuit (see Fig.~\ref{fig.exp}). 
We assume
that a single atomic transition is in resonance with a single mode of
the cavity. To be 
specific, we are particularly interested in an ensemble of $^{87}$Rb atoms
trapped and cooled in a dipole trap in close vicinity (a few $\mu$m) of the
surface of the $LC$-circuit. 
Neutral $^{87}$Rb has a hyperfine split ground state with total
angular momentum
$F=1$ and $F=2$.  The hyperfine splitting
between the states $|F,m_F\rangle$ 
with $F=1$ and $F=2$ is $\omega\simeq 2\pi\times 6.834$\,GHz
\cite{PhysRev.127.524,bize_high-accuracy_1999}. Without an external magnetic
field, the Zeeman sublevels with $z$-component of the total angular momentum  
$m_F=-F,\ldots,F$ are degenerate. We will focus on the situation where an
additional small static $\bB$-field 
in the $z-$direction is applied that splits the degenerate hyperfine
states, and defines the quantization axis for the atoms. Neglecting the magnetic dipole moment of
the nucleus, and for small magnetic field strength $B\ll
\hbar\omega/(g_S\mu_B)$, the magnetic 
moment of an atom is given by  
\begin{equation} \label{muop2}
\bmu=g_S\mu_B\bS/\hbar\,,
\end{equation}
where $g_S\simeq 2$ and $\bS$
are the 
$g$-factor and total electron angular 
momentum vector, respectively, and $\mu_B=e\hbar/(2m_e)$ is the Bohr
magneton (with $e$ and $m_e$ the electron charge and mass). 

We consider a simple current carrying loop with
inductance $L$, that is part of a resonant $LC$ circuit with frequency
$\omega$. The resonator can be treated as a harmonic oscillator with
Hamiltonian $H_{LC}=\hbar\omega(a^\dagger a+1/2)$.  The creation and
annihilation operators $a^\dagger$ and $a$ are related to the
current $I$ in the circuit by $I=\sqrt{\hbar
  \omega/(2L)}(a^\dagger+a)$. That current gives rise to a
magnetic field 
\begin{equation} \label{B}
\bB_{LC}(\bx)=\mu_0\frac{I}{4\pi d}\bb(\bx)\,,
\end{equation}
where $\bb(\bx)$ is a dimensionless mode function with components
$b_x,b_y,b_z$ depending on the
geometry of the $LC$ circuit, $d$ a typical
linear dimension of the $LC$ circuit,  and $\mu_0$ the
magnetic constant.  
The magnetic moment of a single atom at position $\bx$ couples to the
magnetic field through the 
interaction Hamiltonian $H_{\rm int}^{(1)}=-\bmu\cdot \bB(\bx)$, which can
be written in the $|F,m_F\rangle$ basis as 
\begin{equation}
H_{\rm int}^{(1)}=-\left(g_S\frac{\mu_B\mu_0}{4\pi d}\sqrt{\frac{\omega}{2\hbar
  L}}\right)\sum_{F,F',m_F,m_F'}\langle
  F,m_F|\bb(\bx)\cdot\bS|F',m_F'\rangle|F,m_F\rangle\langle
  F',m_F'|(a^\dagger+a)\,.  
\end{equation}
Starting in
$|F,m_F\rangle=|1,0\rangle$, the only 
non-vanishing resonant transition due to a $\bB_{LC}(\bx)$ from the circuit
oscillating at angular frequency $\omega$ is  then to
$|F',m_F'\rangle=|2,0\rangle$ with 
matrix element $\langle 1,0|\bb\cdot\bS|2,0\rangle=\hbar b_z/2$.  Note
that to first order in the additional static $\bB$-field,
$|1,0\rangle$ and $|2,0\rangle$ are not shifted by the 
field. In the interaction picture with respect to the free Hamiltonian
$H_0=H_{LC}+\hbar\omega (|2,0\rangle\langle 2,0|-|1,0\rangle\langle
1,0|)/2$, the operators $a$ and $|2,0\rangle\langle 1,0|$ acquire
time-dependent 
phase factors $e^{-i\omega t}$ and $e^{+i\omega t}$, respectively,
which justifies the use of a
rotating wave approximation.  The interaction Hamiltonian for $N$ atoms
at position $\bx_i$ then takes the familiar form 
\begin{equation}
H_{\rm int}=\sum_{i=1}^N \hbar g_i(a^\dagger \sigma_-^{(i)}+a\sigma_+^{(i)})\,,
\end{equation}
with  $g_i=g(\bx_i)$, 
\begin{equation}
g(\bx)=-\frac{\mu_B\mu_0}{4\pi d}\sqrt{\frac{\omega}{2\hbar L}}b_z(\bx)\,,
\end{equation}
and $\sigma_-^{(i)}=_i|1,0\rangle\langle 2,0|_i$ is the Pauli lowering
operator for atom $i$.

\begin{figure}
(a) \epsfig{file=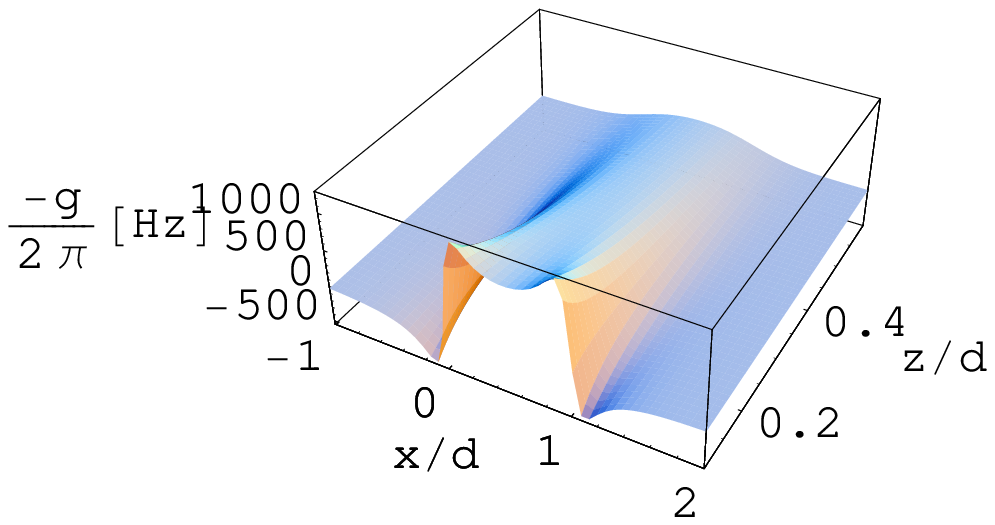,width=7cm,angle=0}
(b) \epsfig{file=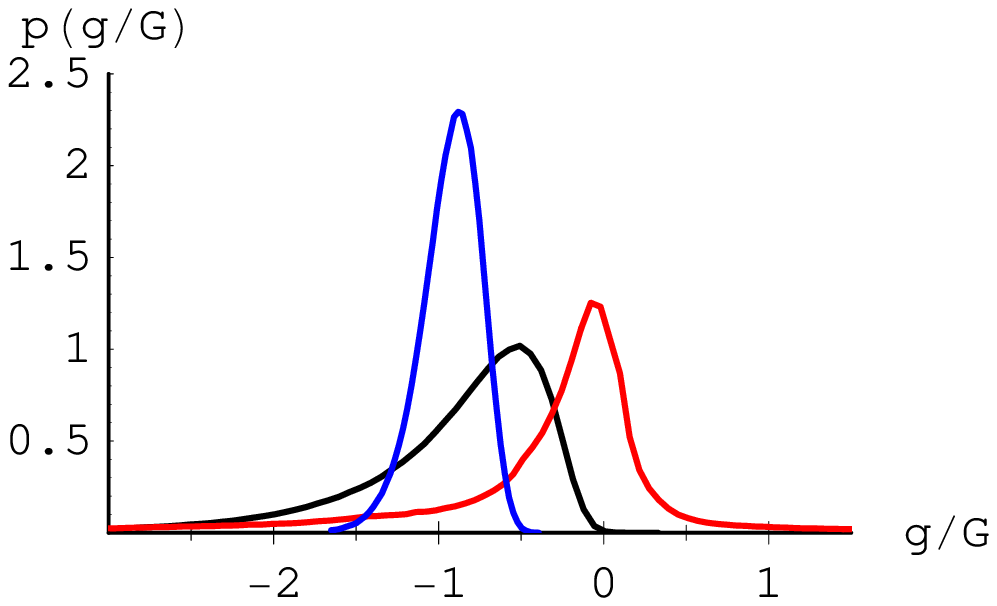,width=7cm,angle=0}
\caption{(Color online) (a) Coupling constant $-g/(2\pi)$ as a
  function of 
  the position $(x,d/2,z)$ for a $^{87}$Rb atom close to a square loop
  $LC$ resonator in the $xy$-plane with
  inductance $L=$1\,pH and 
  resonance frequency $\omega\simeq 2\pi\times 6.834$\,GHz for the
  transition $|F,m_F\rangle=|1,0\rangle\to|2,0\rangle$. (b) Distribution
  of dimensionless coupling constants $g/G$ with
  $G=\frac{\mu_B\mu_0}{4\pi d}\sqrt{\frac{\omega}{2\hbar L}}$ for
  atoms held in a harmonic trap with 
  frequencies $\Omega_x=\Omega_y=\Omega_z=2\pi\times 1$\,kHz centered at
  $(d/2,d/2,d$) with $d=10\mu$m, and otherwise same parameters as in
  (a), and temperatures $T=0.1\,\mu$K (blue/left curve), $T=1\mu\,$K
  (black/middle curve), and $T=10\,\mu$K (red/ right curve).
}\label{fig.coup}        
\end{figure}

In Fig.~\ref{fig.coup} we show $-g_i/(2\pi)$ for an $LC$ circuit in the form
of a 
square loop of size $d=10\mu$m in the $xy$--plane, and a $^{87}$Rb atom
located at 
position $(x,y=d/2,z)$ on the
symmetry axis perpendicular to the loop, where the origin of the coordinate
system is located at the lower left corner of the square, and the center
of the square is $(d/2,d/2,0)$.
The coupling for
an atom at $(d/2,d/2,r)$ with $r=5\,\mu$m is $-g_i\simeq 2\pi\times 388$\,Hz. 
For $r=4\,\mu$m, it increases to   $-g_i\simeq 2\pi\times
504$\,Hz, whereas at $r=6\,\mu$m, we have $-g_i\simeq 2\pi\times 297$\,Hz.
We consider a thermal cloud of atoms confined by a 3D harmonic trap
with trapping frequencies $\Omega_\lambda$ ($\lambda\in \{x,y,z\}$),
centered at $(x_0,y_z,z_0)$ with
atom
positions $\bx_i$ taken as classical variables, distributed according
to the probability density 
$p(\bx)=p_x(x-x_0)p_y(y-y_0)p_z(z-z_0)$. Here,   
\begin{equation}
p_\lambda(\xi)=\frac{1}{\sqrt{\pi}l_{\lambda,T}}\exp(-\xi^2/l_{\lambda,T}^2)
\end{equation}
  is the
thermal equilibrium distribution at 
temperature $T$ for a one-dimensional 
harmonic oscillator of frequency $\Omega_\lambda$ and atomic mass $M$,
with $l_{\lambda,T}=l_\lambda\coth^{1/2}(\hbar
\Omega_\lambda/(2k_BT))$, 
$l_\lambda=\sqrt{\hbar/(M\Omega_\lambda)}$, and $\lambda\in \{x,y,z\}$
(see e.g.~problem 2.6 in \cite{Schleich01}).
From $p(\bx)$ we obtain the distribution of coupling constants 
\begin{equation}
p(g)=\int\,dx dy dz \,\delta(g-g(\bx))p(\bx)\,,
\end{equation}
where the integration is over the entire space $\mathbb{R}^3$.
Examples of $p(g)$ are shown in Fig.~\ref{fig.coup}. 
We see that even at temperatures as low as $0.1\mu$K, relatively
large trapping frequencies of $2\pi\times$1~kHz in all directions, and
a loop size of $d=10\mu$m, the coupling constants can easily vary by 50\% or
more. For higher temperatures $T=1\mu$K, long tails in $p(g)$ with
substantially larger 
absolute coupling constants appear due to atoms close to the loop.
For $T=10\mu$K, the spread in position becomes comparable to $d$, and
many atoms are therefore located ``outside'' the loop, leading to
coupling constants with opposite sign, and a $p(g)$
distributed around an average value close to zero. It is
therefore clear that the consequences of the inhomogeneity of the
coupling constants need to be investigated.

\subsection{Superradiance}
Superradiance is a collective emission process of $N$ two--level atoms
resonantly coupled to a single mode of a resonator (under suitable conditions
it can even be observed without a resonator, see \cite{Gross82}).  It is
observed under 
several 
conditions \cite{Bonifacio71a,Gross82}: 
1.) The temperature of the electromagnetic environment must be much
smaller than the level spacing of the atoms, such that photons only
leave the resonator, whereas the entering of thermal photons can be
neglected, and 2.) the resonator
must be 
rather leaky.  More precisely, if all $N$ atoms couple identically to the
resonator, one should have $\Gamma\ll \sqrt{N}g\ll
\kappa$, where $\kappa$ is the single photon escape rate from the
resonator, and $\Gamma$ denotes the 
single atom spontaneous emission rate. The latter is, in our case of
hyperfine levels of $^{87}$Rb atoms, entirely negligible.   Photon
escape from the 
resonator is due to the finite quality factor, $Q$, of the $LC$ circuit.
$Q$ factors 
of up to $5\cdot 10^5$ have been achieved for superconducting $LC$ circuits
\cite{DiNardo71}, leading to $\kappa=\omega/Q$ of the order of several kHz. For
such a 
high-quality resonator, and the typical coupling constants of a few
100 Hz calculated in section \ref{sec.physsys}, the number of atoms would
be restricted to about 100.  However, the quality factor of the $LC$ circuit
can be easily decreased
 (e.g.~by increasing the
temperature, or using a more lossy substrate), or the
distance of the atoms from the circuit can be increased, thus 
reducing the 
coupling constants.  One can therefore accommodate much larger numbers
of atoms. Below we also derive a lower bound on the  number of atoms
for the validity of the present analysis.

Under the above conditions, and the assumption of weak inhomogeneity
$g_i=g+\delta g_i$ 
with $|\delta g_i|\ll g$, the superradiant
master equation in the rotating frame 
\begin{equation} \label{sr1}
\dot{\rho}=\gamma\left([{\cal  J}_-,\rho {\cal J}_+]+h.c.\right)
\end{equation}
can be derived. Here, $\rho$ represents the
reduced density matrix describing the hyperfine states of the 
atoms, after tracing out the mode of the resonator and its electromagnetic
environment. As mentioned above, the external state of the atoms is
taken as a classical degree of freedom, uncorrelated from the internal
states of the atoms, and with the atoms at fixed positions $\bx_i$ on
the time scale of the experiment.  The rate $\gamma$ is linked to $g$
and $\kappa$ by  $\gamma=g^2/\kappa$.  
The ${\cal J}_\pm$ are collective
ladder operators, related to the single-atom Pauli
matrices in the Hilbert space of the two levels participating in the
resonant transition by
\begin{equation}\label{jpm}
{\cal J}_\pm=\sum\tilde{g}_i\sigma_\pm^{(i)}\,,
\end{equation} 
where $\tilde{g}_i\equiv g_i/g=1+\delta
g_i/g$ denotes a dimensionless coupling  
strength, and $g$ is a reference coupling strength (which might be
chosen e.g.~as the average coupling constant over all atoms). For later use
we also define $\delta\tilde{g_i}=\delta
g_i/g$. 
The derivation of Eq.~(\ref{sr1}) follows closely the calculation in 
\cite{Bonifacio71a}. One checks that in the case of weak asymmetry
the derivation in \cite{Bonifacio71a} remains valid if the
pseudo-angular momentum operators
$J_\pm=\sum_i\sigma_\pm^{(i)}$ used in \cite{Bonifacio71a} are replaced by
${\cal J}_\pm$ defined by Eq.~(\ref{jpm}).

The homogeneous situation, with identical coupling constants, has the tremendous advantage of the dynamics remaining restricted to a single
irreducible 
representation (irrep) of  $SU(2)$ with angular momentum
quantum number 
$j$. In particular, if one starts with all atoms excited, or any other state
fully symmetric under permutation of atoms, then $j=j_{max}=N/2$
is the maximum possible (pseudo-)angular momentum number.  Thus, instead of
having to track $2^N$ states, we only have to deal with $2j_{max}+1=N+1$
states. 
In the following we will consider the situation of weak inhomogeneity
 and assess the effect of the deviations $\delta g_i$
to first order perturbation theory.  
We also introduce a rescaled 
dimensionless time, $\tau=2 J\gamma t$, $J=j_{max}+1/2$, which means
that everything will be expressed in terms of the classical 
time scale of the system \cite{PBraun98b}.  In particular, probabilities
are known to 
propagate on a time scale $\tau\sim 1$, whereas macroscopic coherences in
superradiance 
normally decay on the much shorter time scale $\tau\sim 1/J$. 

Before studying the superradiance dynamics, let us determine the
parameter regime in which we expect our approximation of atoms with
fixed classical positions to be valid. The one-atom reduced density matrix
of a non-interacting atom gas in thermal equilibrium in a 1D harmonic
oscillator in position representation is given by 
\begin{equation}
\langle
x|\rho|x'\rangle=\frac{1}{\sqrt{\pi}l_{x,T}}\exp\left(
-\frac{(x+x')^2}{4l_{x,T}^2}-\frac{(x-x')^2}{4l_{c,T}^2}\right)\,. 
\end{equation} 
We see that the coherences decay on a length scale $l_{c,T}$, which is
given by $l_{c,T}=l_x\tanh^{1/2}(\beta\hbar\Omega_x/2)$ and $\beta=1/k_BT$
\cite{Schleich01}. Quantum effects 
in the external degree of freedom of the atoms can be
neglected if $l_{c,T}\ll l_x$, the harmonic oscillator length scale of
confinement of the 
atoms (and correspondingly for directions $y,z$).  This implies
$\hbar\Omega_x/2\ll k_BT$, which leads to roughly 
$10^{-2}(\nu_x/{\rm kHz})/(T/\mu{\rm K})\ll 1$. Thus, for trapping
frequencies of the order of 1\,kHz treating the atom positions
classically is reasonable down to temperatures of order 100\,nK,
and for trapping frequencies of order 100\,Hz even 10\,nK still
means that the atom positions are essentially classical variables. 

A second restriction arises from the requirement of fixed atom positions
on the time scale of the experiment.  The typical thermal
velocity in the $x$-direction of an atom follows from the equipartition
theorem, $v_x=\sqrt{k_B T/M}$. A typical time scale of an experiment
is given by the decay time of the superradiant pulse, $t\simeq
1/(N\gamma)$. During that time the atom should move at most a distance
much smaller than $d$, the typical length-scale on which $g(\bx)$
changes. This leads to a lower bound on the number of atoms, $N\gg
\sqrt{k_B T/M}/(\gamma d)$. Together with the upper bound $N\ll
(\kappa/g)^2$ for the existence of superradiance, one obtains an
allowed range of permissible values of $N$.  Requesting an upper bound
much larger than the lower bound leads to the requirement $T\ll
M(\kappa d)^2/k_B\simeq (\kappa/{\rm MHz})^2(d/\mu 
{\rm m})^2$ $\times$ 8\,mK in the case of $^{87}$Rb. This leaves a comfortable
range of the Rb temperatures for resonators with a quality of order 1000,
corresponding to $\kappa\simeq 1$\,MHz. 
Using temperatures close to  the lower bound considered above will
have the advantage that one may even record many runs with 
the same atom positions, and thus separate quantum fluctuations
from the superradiant dynamics (e.g.~waiting time distributions after
initial excitation) from fluctuations due to changing coupling
constants. As an example, for $\kappa\simeq 10^6$\,Hz,
$g\simeq 400$\,Hz, we have the requirement $N\ll 10^7$.  At a
temperature of $1\,\mu $K, 
the thermal speed of the Rb atoms is of the order 1\,cm/s, and for
$d=10$\,$\mu$m, we need $N\gg 20$.
  
We now proceed with the analysis by rewriting
(\ref{sr1}) as 
\begin{eqnarray} \label{sr2}
\frac{d\rho}{d\tau}&=&\cL[\rho]\\
\cL[\rho]&=&\cL_0[\rho]+\epsilon\cL_1[\rho]\\
\cL_0[\rho]&=&\frac{1}{2J}\left([J_-,\rho J_+]+h.c.\right)\\
\cL_1[\rho]&=&\frac{1}{2J}\sum_{i,j}\left(\delta
\tilde{g}_i+\delta\tilde{g}_j^*+\delta\tilde{g}_i\delta\tilde{g}_j^*\right) 
\left([\sigma_-^{(i)},\rho\sigma_+^{(j)}]+h.c.\right) \\
&\simeq&\frac{1}{2J}\sum_{i,j}\left(\delta
\tilde{g}_i+\delta\tilde{g}_j^*\right)
\left([\sigma_-^{(i)},\rho\sigma_+^{(j)}]+h.c.\right) \,,\label{L1}
\end{eqnarray}
where in the last step we have neglected the terms of order
$\delta\tilde{g}^2$. We have introduced a book keeping parameter
$\epsilon$, assuming that $\cL_1$ is small compared to
$\cL_0$. In the end we will set $\epsilon=1$.   
We 
also expand  $\rho$ in terms of $\epsilon$,
\begin{equation} \label{rhoe}
\rho=\rho_0+\epsilon \rho_1+\epsilon^2 \rho_2\ldots\,.
\end{equation}
To order $\epsilon^0$, we retrieve the original Lindblad master equation
\begin{equation} \label{L0}
\frac{d\rho_0}{d\tau}=\cL_0[\rho_0]
\end{equation}
with initial condition $\rho_0(0)=\rho(0)$. To first order, $\epsilon^1$, we
get 
\begin{equation} \label{e1}
\frac{d\rho_1}{d\tau}=\cL_0[\rho_1]+\cL_1[\rho_0]\,,\,\,\,\rho_1(0)=0\,.
\end{equation}
The formal solution of (\ref{e1}) is given by 
\begin{equation} \label{rho1}
\rho_1(\tau)=\int_0^\tau d\tau'
e^{\cL_0(\tau-\tau')}\cL_1e^{\cL_0\tau'}[\rho(0)]\,. 
\end{equation}
It describes the usual situation of propagating the initial density
matrix with the ``free'' propagator $\exp(\cL_0 \tau')$ corresponding to the
homogeneous case up to a time $\tau'$, then have the perturbation
$\cL_1$ act at time $\tau'$, and then continue with the ``free'' propagation
till the end of the time interval. The free propagator has been well studied,
and in particular very precise semi-classical expressions exist, both
for 
the propagation of the probabilities and for the coherences --- but
only for $j=j_{max}$ relevant for the homogeneous case
\cite{Agarwal70,Bonifacio71a,Bonifacio71b,Glauber76,PBraun98a,PBraun98b}.
Here, we have the added complication that 
$\cL_1$ does not conserve $j$, as $\cL_1$ is by
definition {\em not} symmetric under exchange of atoms. We therefore
need a more general expression for the free propagator for the time interval
$[\tau',\tau]$. We will derive
here a free propagator valid in the entire 
$2^N$ dimensional Hilbert space, represented in all $SU(2)$ irrep
components, by generalizing the method in \cite{PBraun98a} that allows
one to
connect the propagator for coherences to the one for
probabilities. Secondly, we will obtain explicit
expressions for the matrix elements of $\cL_1$ in and between different
$SU(2)$ irreps. Taken together, this will allow the construction of
the full propagator according to Eq.~(\ref{e1}).

The irreducible representations of $SU(2)$ can be constructed by
adding one spin-1/2 after 
another. When adding a new spin, the available values of total angular
momentum $j$,
can either increase or decrease by $1/2$.  Therefore, for $N>2$, there are many ways to get to a particular value $j$ for
given number of atoms $N$. These different ways lead to different
degenerate irreps for 
the same $j$. A quantum state must accordingly be labeled not only by the
total angular momentum $j$ and $m$ ($m=-j,\ldots,j$), but also by additional
quantum numbers $\alpha$, which distinguish different irreps with the
same $j$. We may
consider $\alpha$ as the label of a path on the lattice of allowed
$(j,N)$ combinations  that leads to the value of $j$ at hand.  We thus have
states $|\alpha;j m\rangle$ that are eigenstates of $\bJ^2$ and $J_z$ with
eigenvalues $j(j+1)$ and $m$, respectively, which we can use to represent
$\cL_1$. It will turn out that to first order 
perturbation theory, one needs only the irreps with $j=j_{max},j_{max}-1$
and $j_{max}-2$. This is a consequence of the fact that $\cL_1$ only contains 
tensor operators of rank 1 and 2.

\section{Superradiance for homogeneous couplings in the entire Hilbert space}
It can be shown that the superradiant master equation (\ref{sr1}) with
homogeneous 
couplings conserves both $j$ and
$\alpha$, i.e.~
\begin{equation}
\langle
\beta;j'm'|J_{\pm}|\alpha;jm\rangle=\delta_{j'j}
\delta_{\alpha\beta}\delta_{m'\,m\pm 
  1} d_\pm(j,m)
\end{equation}
with $d_\pm(j,m)\equiv \sqrt{j(j+1)-m(m\pm 1)}$.
 Conservation of $j$ has been known from the early days of superradiance and
 allowed a
formulation of the problem in the $j=j_{max}$ irrep.
Conservation of $\alpha$ can be shown by complete induction.  
Below we will consider the  situation of two
sub-ensembles of atoms, each of which is homogeneous on its own.  In that
case we 
will see that conservation of $\alpha$ also follows from selection
rules encoded in the angular momentum algebra.

It is worthwhile to express the matrix elements of $\rho$
as  
\begin{equation} \label{rhom}
\rho_m(j,j',k,\alpha,\alpha',\tau)\equiv\langle \alpha;j
\,m+k|\rho(\tau)|\alpha';j'\,m-k\rangle\,,
\end{equation}
where $m$ is a ``center of mass'' quantum number,
$m=(m_1+m_2)/2$, and $k=(m_1-m_2)/2$ for a matrix element
$\bra{\alpha;j,m_1}\rho\ket{\alpha';j',m_2}$.
A value $k$ different from zero, $j\ne j'$, or $\alpha\ne \alpha'$
signifies a coherence. The quantum numbers   $m$ and $k$ are simultaneously
integer or half-integer.  In order to render the notation
less cumbersome, we denote the set of quantum numbers
$(j,j',k)$ collectively as $x$.
We can then  write Eq.~(\ref{L0}) as 
\begin{equation} \label{sr3}
\frac{d}{d\tau}\rho_m(x,\alpha,\alpha',\tau)=a_{m+1}(x)\rho_{m+1} (x,\alpha,\alpha',\tau)-b_m(x)\rho_m(x,\alpha,\alpha',\tau)\,,
\end{equation}
 and see that probabilities and coherences do not mix under time evolution,
 nor do coherences defined through different combinations of
 $j,j',\alpha,\alpha'$, and $k$. The coefficients $a_m$ and $b_m$ are
 independent of $\alpha,\alpha'$, and are defined as 
\begin{eqnarray}
a_m(x)\equiv a_m(j,j',k)&=&\frac{1}{J}d_-(j,m+k)d_-(j',m-k)\nonumber \\
b_m(x)\equiv b_m(j,j',k)&=&\frac{1}{2J}\left(d_+(j,m+k-1)d_-(j,m+k)+d_-(j',m-k)d_+(j',m-k-1)\right)\,.\nonumber
\end{eqnarray}
The master equation can be
solved with the help of the propagator $D=\exp(\cL_0 \tau)$, which is
non-diagonal now only in the quantum numbers $m$.  We can thus write
\begin{equation} \label{rm}
\rho_m(x,\alpha,\alpha',\tau)=\sum_{n=n_\downarrow(x)}^{n_\uparrow(x)}D_{mn}(x,\tau)\rho_n(x,\alpha,\alpha',0)\,.
\end{equation}
where we have used that the propagator $D$ does not depend on
$\alpha,\alpha'$. 
The sum over $n$ runs over all values from 
$n_\downarrow(x)=\max(-j-k,-j'+k)$ to
$n_\uparrow(x)=\min(j-k,j'+k)$ in steps of 1, and $m$ is
restricted to the same interval. These bounds follow from requesting $-j\le 
m+k\le j$ and $-j'\le m-k\le j'$.
The initial condition is
$D_{mn}(x,0)=\delta_{mn}$. The equation can be solved exactly by
Laplace transformation, 
\begin{equation} \label{lp}
\tilde{D}_{mn}(x,z)=\int_0^\infty e^{-z\tau}D_{mn}(x,\tau)d\tau\,,
\end{equation}
which leads to the recursion relation
\begin{equation} \label{rec}
(z+b_m(x))\tilde{D}_{mn}(x,z)=a_{m+1}(x)\tilde{D}_{m+1\,n}(x,z)+\delta_{mn}\,.
\end{equation}
The solution
\begin{equation} \label{dmn}
\tilde{D}_{mn}(x,z)=\frac{1}{a_m(x)}\prod_{l=m}^n\frac{a_l(x)}{z+b_l(x)}\, 
\end{equation}
valid for all $m\le n$, has poles at $z=-b_l(x)$. 
It remains to perform the inverse Laplace transform,
\begin{equation} \label{invlp}
D_{mn}(x,\tau)=\frac{1}{2\pi i}\int_{z_0-i\infty}^{z_0+i\infty}dz\,e^{z\tau}\tilde{D}_{mn}(x,z)\,,
\end{equation}
where $z_0\in\mathbb{R}$ needs to be chosen larger than the real part of all
poles. Unfortunately, for 
large $N$, there are many poles, which makes the inverse Laplace
transform cumbersome.  However, we can relate the propagator for
coherences to the propagator
of probabilities \cite{PBraun98a} by noting that
\begin{eqnarray}
b_l(x)&=&b_l(j,j,0)-\left(j(j+1)-j'(j'+1)+2k^2\right)/(2J)\\
a_l(x)&=&a_l(j,j,0)q_l(x)\,,\\
q_l(x)&=&\frac{\sqrt{(j-k-l+1)(j+k+l)(j'+k-l+1)(j'-k+l)}}{(j+l)(j-l+1)}\,.  
\end{eqnarray}
Once we shift the
integration variable $z$ in Eq.~(\ref{invlp}) by
$-\left(j(j+1)-j'(j'+1)+2k^2\right)/(2J)$ and 
define
\begin{equation} \label{Q}
Q(x,m,n)=\prod_{l=m+1}^n q_l(x)\,,
\end{equation}
we find
\begin{equation} \label{cohpro}
D_{mn}(x,\tau)=e^{(j(j+1)-j'(j'+1)+2k^2)\tau/(2J)}
Q(x,m,n)D_{mn}(j,j,0,\tau)\,,   
\end{equation}
where $D_{mn}(j,j,0,\tau)$ is the propagator of probabilities.
A precise semiclassical approximation for $D_{mn}(j,j,0,\tau)$
can be found 
in \cite{PBraun98a,PBraun98b}. We thus have through Eq.~(\ref{cohpro})
immediate access to a propagator in the entire exponentially large Hilbert
space. 

The independence of 
$D_{mn}(j,j',k,\tau)$ on $\alpha,\alpha'$ implies that coherences
$\rho_m(x,\alpha,\alpha',\tau)$ with 
$k=0$ 
between irreps with different $\alpha,\alpha'$, but the same $j$, decay just
as slowly as the probabilities within any irrep with the same $j$. This
constitutes another example of slowly decohering Schr\"odinger cat states,
with which the theory of superradiance is rich \cite{Braun98}.  The
physical reason for this 
is that the Lindblad operator $\cL_0[\rho]$ is totally symmetric under
permutation of 
atoms, and therefore cannot make transitions between irreps with different
values of $\alpha$.  Previously known examples of slowly decohering
Schr\"odinger cat states in superradiance include superpositions of
angular momentum coherent states in the $j=j_{max}$ irrep that are
symmetric with respect to the 
equator $J_z=0$. They also decohere just as slowly as the corresponding
probabilities.  Furthermore, there is a well known decoherence
free subspace (DFS) of dimension 
${N\choose N/2}\sim 2^N/\sqrt{N}$ that contains the states
$|\alpha;j,-j\rangle$ of all
irreps $(\alpha,j)$ in which all superpositions, as macroscopic or
entangled as 
they may be, are 
decoherence free under this collective decoherence process
\cite{Braun01B}.

\section{Two subsystems}
The superradiance process depends on all individual coupling constants.  To
simplify the problem at hand, let us look at only two sub-ensembles with
$N_1$ and $N_2$ atoms ($N_1+N_2=N$), and coupling constants $g_1$ and $g_2$,
respectively. While this exact distribution of coupling constants may
not be very realistic, one may hope to develop an idea of the
effects of inhomogeneity in this situation that may still be
qualitatively correct for more complicated distributions of
couplings, and to find a solution that is still analytically tractable. 
For an arbitrary distribution of coupling
constants one may choose to adjust $N_1$, $g_1$ and $g_2$ to the first three
nontrivial moments of the distribution.  The results are
\begin{equation}
g_{1,2}=\frac{{\dgthree} - {\dgtwo}\,{\langle g\rangle} \pm 
    {\sqrt{4\,{{\dgtwo}}^3 + {{\dgthree}}^2 - 
        6\,{\dgtwo}\,{\dgthree}\,
         {\langle g \rangle} + 
        9\,{{\dgtwo}}^2\,{{\langle g\rangle}}^2}}}{2\,
    {\dgtwo}}\,,
\end{equation}
and 
\begin{equation}
N_1-N_2= N\frac{\dgthree-3 \dgtwo\, \langle g\rangle}{\sqrt{4
    \dgtwo^3+\dgthree^2-6 \dgtwo\dgthree  
    \langle g\rangle+\dgtwo^2\langle g\rangle^2}}\,. 
\end{equation}
Here,
$\dgtwo\equiv\sum_{i=1}^N(g_i-\langle g\rangle)^2/N$, etc. We assume that 
$N$ is big enough such that rounding to the next integer or half integer
value for $N_1$ does  
not lead to a significant change. 
The advantage of working with just two different couplings is that
explicit analytical formulas for $\cL_1$ can be obtained. We define
the eigenstates $|j_i,m\rangle$ of $\bJ^{(i)}.\bJ^{(i)}$ and
$J_z^{(i)}$ for $i=1,2$.  The basis
functions $\ket{(j_1,j_2)jm}$ 
which are eigenstates of $\bJ^{(1)}.\bJ^{(1)}$, $\bJ^{(2)}.\bJ^{(2)}$,
$\bJ.\bJ$, and $J_z$, will be used to represent the state of the whole
system. The symmetry
label $\alpha$ corresponds then to the 
pseudo-angular momentum quantum numbers $j_1$, $j_2$ for the two
sub-ensembles. 

For given couplings $g_1=g+\delta g$ and $g_2=g-\delta g$,
we have $\delta g=(g_1-g_2)/2$. We also introduce
the ladder 
operators 
$\Jp{1}$, $\Jm{1}$, $\Jp{2}$, and $\Jm{2}$ in the two subsystems, defined as
\begin{equation} \label{j1pm}
J_\pm^{(1)}=\sum_{i=1}^{N_1}\sigma_\pm^{(i)},\,\,\,\,J_\pm^{(2)}=\sum_{i=N_1+1}^{N}\sigma_\pm^{(i)}\,.
\end{equation}
$\cL_1$ in Eq.~(\ref{L1}) then becomes
\begin{eqnarray}
\cL_1[\rho]&=&\frac{\delta\tilde{g}}{J}\left(2(\Jm{1}\rho\Jp{1}-\Jm{2}\rho\Jp{2})-\rho(\Jp{1}\Jm{1}-\Jp{2}\Jm{2})-(\Jp{1}\Jm{1}-\Jp{2}\Jm{2})\rho\right)\,.
\end{eqnarray}
We then need to calculate the matrix elements of $\Jm{1}$,$\Jm{2}$,
and $\Jp{1}\Jm{1}$, 
$\Jp{2}\Jm{2}$ in basis states $\ket{(j_1,j_2)jm}$. The results can be found
in the Appendix.
All these matrix elements are real.  The
expressions for $\Jm{2}$ and $\Jp{2}\Jm{2}$ are obtained from those for
$\Jm{1}$ and $\Jp{1}\Jm{1}$ 
simply by exchanging $j_1\leftrightarrow j_2$, which evidently leaves the
expressions unchanged if $j_1=j_2$. Equation (\ref{L1}) then tells
us that $\cL_1[\rho]=0$ for $j_1=j_2$, for any $\rho$ that is
symmetric under the exchange of the two subsystems, $1\rightarrow
2$. This means that to first order 
perturbation theory the effect of the inhomogeneity in the couplings vanishes
exactly if the two sub-ensembles contain the same number of atoms and the
initial state is in the $j=j_{max}$ irrep.

\subsection{Full propagator}
Let us consider the case of an initially pure state,
$\rho(0)=|\psi(0)\rangle\langle\psi(0)|$ with support only in the irrep with
$j=j_{max}$. Since
we start off with a totally symmetric state, the initial 
state is also totally symmetric under permutations in each sub-ensemble, and
we have therefore 
initially $j_1=N_1/2$, $j_2=N_2/2$. To zeroth order we find
\begin{eqnarray}
\rho_0(\tau)_{jm,j'm'}&=&\delta_{j,j'}\delta_{j,j_{max}}\sum_n
D_{\frac{m+m'}{2}n}(j,j,\frac{m-m'}{2},\tau)\psi^*_{n+\frac{m-m'}{2}}(0)\psi_{n-\frac{m-m'}{2}}(0)\,, 
\end{eqnarray}
with $\psi_m(0)\equiv \langle(j_1,j_2)j_{max}m\ket{\psi(0)}$.
The first order correction to the density matrix  can be written as 
\begin{eqnarray}
\rho_1(\tau)_{jm,j'm'}&=&\delta\tilde{g}\sum_{s,r,r'}P_{jmj'm';srr'}
\psi^*_{s+\frac{r-r'}{2}}(0) \psi_{s-\frac{r-r'}{2}}(0)\,,
\end{eqnarray}
with a propagator $P$ defined as
\begin{eqnarray}
P_{jmj'm';srr'}&=&\int_0^\tau\sum_{n}D_{\frac{m+m'}{2}n}(j,j',\frac{m-m'}{2},\tau-\tau')D_{\frac{r+r'}{2}s}(j_{max},j_{max},\frac{r-r'}{2},\tau')d\tau'\nonumber\\ 
&&\cLL{j}{j_{max}}{n+\frac{m-m'}{2}}{r}{j'}{j_{max}}{n-\frac{m-m'}{2}}{r'}\,.\label{prop1}
\end{eqnarray}
The sums over $s,r,r'$ are restricted such that $-j_{max}\le
s+(r-r')/2,s-(r-r')/2\le j_{max}$. We have introduced the
representation of $\cL_1$ in the $SU(2)$ irrep states,
\begin{eqnarray}
\cLL{j}{l}{m}{r}{j'}{l'}{m'}{r'}&=&\frac{1}{j_{max}+\frac{1}{2}}\Big(2\bra{jm}\Jm{1}\ket{lr}\bra{j'm'}\Jm{1}\ket{l'r'}^*\nonumber\\ 
&&-\delta_{jl}\delta_{mr}\bra{l'r'}\Jp{1}\Jm{1}\ket{j'm'}-\delta_{j'l'}\delta_{m'r'}\bra{jm}\Jp{1}\Jm{1}\ket{lr}-(j_1\leftrightarrow j_2)
\Big)
\end{eqnarray}
With these expressions we can now evaluate any time dependent expectation
value. 

\section{Results}
\subsection{Short time behavior of population inversion}
The population inversion is given by 
$J_z=\frac{1}{2}\sum_{i=1}^N\sigma^{(i)}_z$. To zeroth order we have
\begin{eqnarray} 
\langle
J_{z,0}(\tau)\rangle&=&\tr\left(J_z\rho_0(\tau)\right)=\sum_{m=-j_{max}}^{j_{max}}m
\rho_0(\tau)_{j_{max}m,j_{max}m}\nonumber\\
&=&\sum_{mn}m D_{mn}(j_{max},j_{max},0,\tau)|\psi_n(0)|^2
\,,
\end{eqnarray}
 and to first order
\begin{eqnarray} 
\langle
J_{z,1}(\tau)\rangle&=&\delta\tilde{g}\sum_{j,m,s,r,r'}m P_{jmjm;srr'}(\tau)
\psi^*_{s+\frac{r-r'}{2}}(0) \psi_{s-\frac{r-r'}{2}}(0)\,.\label{Jz1}
\end{eqnarray}
The propagator involves the matrix elements
$\cLL{j}{l}{m}{r}{j'}{l'}{m'}{r'}$ now 
only for $j'=j,m'=m$, and $l=l'=j_{max}$.  Since $j\le j_{max}$, the
summation over $j$ in (\ref{Jz1}) is restricted to
$j=j_{max},j_{max}-1,j_{max}-2$ (see the remarks after
Eq.~(\ref{Jp12})). Moreover, it is possible to obtain closed, 
compact expressions for these matrix elements, at least in the case of
$r=r'$, which is relevant if the initial state is a Dicke state
$|(j_1,j_2)j_{max}m\rangle$. To this end it
is useful to  
consider the three cases $j=j_{max},j_{max}-1$ and $j=j_{max}-2$
separately.
\begin{enumerate}
  \item For $j=j_{max}$, matrix elements of $\Jm{1}$ and of $\Jp{1}\Jm{1}$
  contribute, and we find 
\begin{eqnarray}
\cLL{j_{max}}{j_{max}}{n}{r}{j_{max}}{j_{max}}{n}{r}&=&2\frac{2j_1-j_{max}}{j_{max}(j_{max}+1/2)}\Big[\delta_{n,r-1}(j_{max}+j_{max}^2-n-n^2)\nonumber\\
&&-\delta_{n,r}(j_{max}+j_{max}^2+n-n^2)\Big]\,.\label{jmaxcont}
\end{eqnarray}
\item For $j=j_{max}-1$, only $\Jm{1}$ contributes, due to the
  Kronecker-deltas that come with $\Jp{1}\Jm{1}$. We obtain
\begin{eqnarray}
\cLL{j_{max}-1}{j_{max}}{n}{r}{j_{max}-1}{j_{max}}{n}{r}&=&2\delta_{n,r-1}\Big[\frac{j_1(j_{max}-j_1)(-1+j_{max}+n)(j_{max}+n)}{(j_{max}+1/2)j_{max}(2j_{max}-1)}-(j_1\leftrightarrow
    j_2)\Big]\label{j-1}\nonumber\\
&=&0\,,
\end{eqnarray}
where in the last step we have used $j_2=j_{max}-j_1$ such that the prefactor
$j_1(j_{max}-j_1)=j_1j_2$ is 
symmetric under $j_1\leftrightarrow j_2$.
\item For $j=j_{max}-2$, we find immediately that this contribution
  vanishes, as $\Jm{1}$ cannot change $j_{max}$ by more than one unit, and
  for the $\Jp{1}\Jm{1}$ term the prefactor is again zero due to the
  Kronecker-deltas. 
\end{enumerate}
The only contribution to $\langle J_{z,1}\rangle$ stems
therefore from the original fully symmetric irrep with
$j=j_{max}$. Experimentally, the most relevant situation is a fully excited
initial state 
$|(j_1,j_2)j_{max},j_{max}\rangle$, which can be achieved e.g.~by optical
pumping with external laser light that acts on all atoms in the same way.
In this case, one obtains the explicit result for
the first order correction, 
\begin{eqnarray}
\langle
J_{z,1}(\tau)\rangle&=&2\delta\tilde{g}\frac{2j_1-j_{max}}{j_{max}(j_{max}+1/2)}
\int_0^\tau\sum_{n,m}mD_{mn}(j_{max},j_{max},0,\tau-\tau')\nonumber\\
&&\times\Big[(j_{max}+j_{max}^2-n-n^2)D_{n+1,j_{max}}(j_{max},j_{max},0,\tau')\nonumber\\
&&-(j_{max}+j_{max}^2+n-n^2)D_{n,j_{max}}(j_{max},j_{max},0,\tau')\Big]d\tau'\label{jz1fin} \,.
\end{eqnarray}
Equation (\ref{jz1fin}) is one of the main results of this
paper. It
shows once more that $\langle 
J_{z,1}(\tau)\rangle$ vanishes for $j_1=j_2=j_{max}/2$. Moreover, all
dependence on $N_1$ is in the prefactor $2j_1-j_{max}=N_1-N/2$, such that for
given $N$ it is sufficient to calculate $\langle
J_{z,1}(\tau)\rangle$ for any $N_1\ne N/2$, and then rescale accordingly. 
\begin{figure}
\epsfig{file=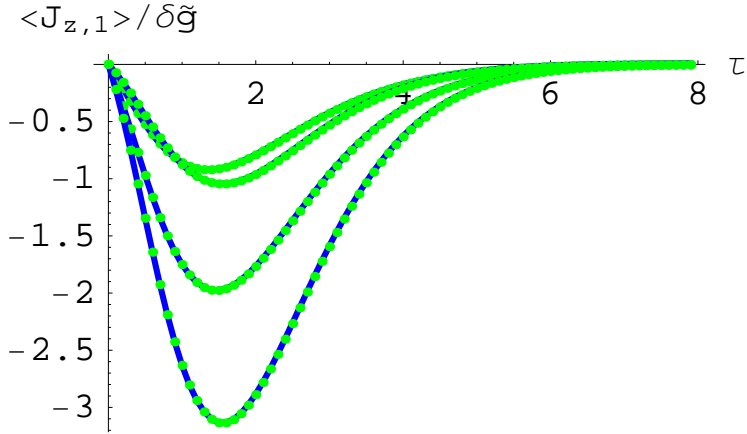,width=10cm,angle=0}
\caption{(Color online) First order correction $\langle
J_{z,1}(\tau)\rangle/\delta\tilde{g}$ to the population inversion
$\langle 
J_{z}(\tau)\rangle$ as function of dimensionless time $\tau$. Green dots are
obtained from the perturbation 
calculation, 
Eq.~(\ref{Jz1}); blue continuous line: exact result from numerical
diagonalization of full propagator for a small number of atoms. Curves
from top to bottom correspond to 
$(N_1,N_2)=(2,1)$, (3,2), (3,1), and (4,1), respectively.  
}\label{fig.jz1}        
\end{figure}
Figure \ref{fig.jz1} shows  $\langle J_{z,1}(\tau)\rangle$
calculated from Eq.~(\ref{Jz1}) and from exact
diagonalization of the 
full propagator for a small number of atoms and for very small
$\delta\tilde{g}$, such that $\langle
J_{z,1}(\tau)\rangle/\delta\tilde{g}\simeq (\langle
J_{z}(\tau)\rangle|_{\delta\tilde{g}}-\langle
J_{z}(\tau)\rangle|_{\delta\tilde{g}=0})/\delta\tilde{g}$. The agreement is
perfect.   
We 
see that $\langle J_{z,1}(\tau)\rangle$ vanishes at $\tau=0$, and then
decreases as function of 
time, before increasing again.  This implies a more rapid initial decay of
the population inversion 
for $\delta \tilde{g}>0$  than for the homogeneous case. As the loss of
atomic excitation goes along with an increase of the photon number in the
cavity, superradiance is  
accelerated by the inhomogeneity. 
Note that $\delta \tilde{g}>0$ together with $j_1>j_2$ means that 
there are more atoms with the larger of the two coupling constants,
such that one indeed expects an acceleration of superradiance compared to
the homogeneous case. For $j_1<j_2$, $\langle J_{z,1}(\tau)\rangle$ changes
sign, and superradiance slows down.  

Figure~\ref{fig.jz1} also shows that for sufficiently large
times, $\langle J_{z,1}\rangle$ vanishes.  This can be understood
from Eq.~(\ref{jz1fin}), as for large $\tau$ the two free propagators under
the 
integral cannot be simultaneously substantially different from zero for the
combinations of indices that appear.  As a consequence, the effect of
inhomogeneity on the 
average value of $\langle J_z(\tau) \rangle$ for $\tau\gg 1$ is at
most  a quadratic
function of $\delta\tilde{g}$, even for $j_1\ne
j_2$  (see also next subsection).  This means that there is a finite time
after which first order perturbation theory becomes inadequate, and
higher order 
corrections will dominate.

The time-independent
factors in Eq.~(\ref{jz1fin}) scale proportional to $j_{max}^2$ for
$j_{max}\gg 1$. However, the
integration over $\tau$  brings down a factor 
$1/(j_{max}+1/2)$ (see
Eq.~(\ref{cohpro}), and Eq.~(4.56) in \cite{Braun01B}). The free propagators
are of order 1, such that 
$\langle
J_{z,1}(\tau)\rangle/j_{max}$ 
scales as $\delta g j_{max}^0$  as it should, and first
order perturbation theory remains   
meaningful for large $j_{max}$. This means that it is enough to
calculate $\langle
J_{z,1}(\tau)\rangle/j_{max}$ for moderate values of $j_{max}$, as it
will saturate as function of $j_{max}$. 
 
\subsection{Incomplete Relaxation}
Another important consequence of inhomogeneous coupling constants that can be 
observed even for a very small number of atoms is incomplete relaxation
(neglecting spontaneous emission --- see the remarks in
sec.\ref{sec.physsys}).  It is well known that for fully $SU(2)$-symmetric
superradiance there is a large decoherence free subspace (DFS)
\cite{Braun01B,Beige00b} containing $N \choose N/2$ dark states for $N$ even
(of $N \choose (N-1)/2$ for $N$ odd) . These states, defined through
$J_-|\psi\rangle=0$, can trap the dynamics, in the sense that if such a state
is reached, the superradiant dynamics is switched off and further evolution
is only possible through competing mechanisms neglected so far.   The
simplest example is given for $N=2$ with two DFS states.  If we denote
the two hyperfine 
states involved as $|0\rangle=|F,m_F=1,0\rangle$ and
$|1\rangle=|F,m_F=2,0\rangle$, the DFS states are
the ground state $|00\rangle$ and the ``singlet'' state
$(|01\rangle-|10\rangle)/\sqrt{2}$. In the singlet state both atoms together
contain one photon, but destructive interference prevents the
transfer of the photon from the atoms to the cavity (from where it would
escape). A way of reaching that state is to start in an initial state
$|01\rangle$, which in half the cases will emit a photon, but in the other
half get trapped in the singlet state \cite{Plenio99}.  This constitutes a
simple way 
of preparing an entangled state through a decoherence (and even dissipation)
mechanism: If no photon leaves within a time given by max( 1/$g$, 1/ $\kappa$),   
the system is with high probability in the singlet state. 

More generally, for perfect $SU(2)$ symmetry, the DFS states are the
(typically highly 
degenerate) ground
states of all the $SU(2)$ irreps with $j=0,1,\ldots, N/2$ (assuming $N$
even).  If the $SU(2)$ symmetry is broken, the DFS does not disappear,
but rather gets rotated in Hilbert space.  For example, for 
$N=2$, and real coupling constants $g_1$ and $g_2$, the singlet is replaced
by a 
state
$g_2|0\rangle_1|1\rangle_2-g_1|1\rangle_1|0\rangle_2/\sqrt{g_1^2+g_2^2}$,
which is still annihilated by
$g_1\sigma_-^{(1)}+g_2\sigma_-^{(2)}$, the new collective
Lindblad-operator.  This parametric dependence of the DFS on a system
parameter is at the basis of ``decoherence-enhanced measurements''
\cite{Braun09,Braun10.2}, which allow precision measurements
with Heisenberg-limited sensitivity while using initial product states.

With perfect $SU(2)$ symmetry, an initially fully excited state with
$j=N/2=j_{max}$ remains in that 
irrep
and relaxes to the ground state, without ever reaching a nontrivial
DFS state, which only exists for $j\ne j_{max}$. However, when the
$SU(2)$ symmetry is broken, $j$ is no longer conserved, and nontrivial DFS
states can be reached, resulting in the trapping of the population.  
Since the first order correction to $\langle J_z(\tau)\rangle$
vanishes for large $\tau$, the trapping effect is beyond reach of the
perturbation theory  developed above. We therefore resort to
a numerical approach by simulating the stochastic Schr\"odinger equation (SSE)
that unravels 
Eq.~(\ref{sr2}). The SSE is given by  
\begin{eqnarray} 
d\psi(t)&=&D_1(\psi(t))\,dt+D_2(\psi(t))\,dW(t)\,,\label{SSE}\\
D_1(\psi)&=&\gamma\left(2\langle J_-\rangle_\psi J_--J_+J_--\langle
  J_-\rangle^2_\psi\right)\psi\\
D_2(\psi)&=&\sqrt{2\gamma}\left(J_--\langle J_-\rangle_\psi\right)\psi\,,
\end{eqnarray} 
$dW(t)$ is a Wiener process with average zero and variance
$dt$, and $\langle J_-\rangle_\psi=\langle \psi|J_-|\psi\rangle$
\cite{Breuer06}. 
When averaging over a large number of realizations of the stochastic
process one obtains a numerically exact solution of the master
equation.  In principle this can be done even for arbitrary coupling
constants, but we stay with the situation of two different
subsystems.  A drawback of the numerical approach is that it is
limited to a small number of atoms.  

\begin{figure}
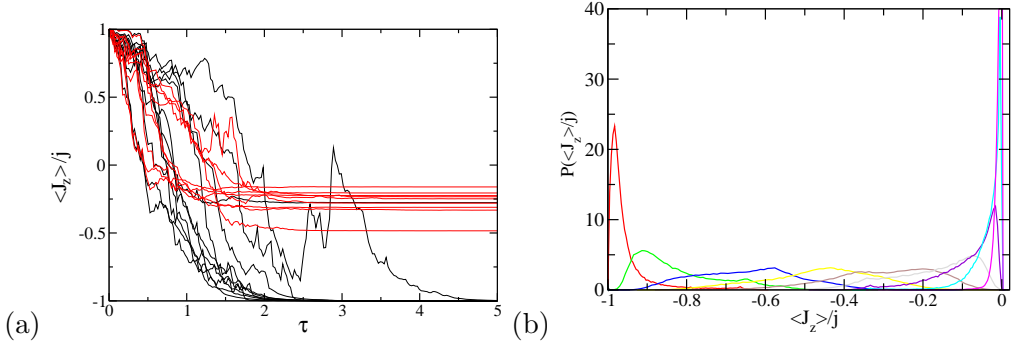

(a) \epsfig{file=jzoftau_random_v2.eps,width=6cm,angle=0}
(b) \epsfig{file=PofJzN6nr1d5_v2.eps,width=6cm,angle=0}
\caption{(Color online) (a) Ten different realizations of the stochastic
  Schr\"odinger equation (\ref{SSE}) for $\delta\tilde{g}=0$ (black) and $\delta\tilde{g}=0.5$
  (red/grey) for $N=12$, $N_1=N_2=6$. Plotted is $\langle \psi(\tau)| 
J_z|\psi(\tau)\rangle/j$ for randomly evolving states $|\psi(\tau)\rangle$.
  (b) Histogram  of $\langle   
J_z(\tau)\rangle/j$ for large $\tau$ ($\tau=5$) for $N=6$, $N_1=N_2=3$
  from $10^5$ realizations of the 
SSE Eq.~(\ref{SSE}) for different values of $\delta\tilde{g}$ 
($\delta\tilde{g}=0.1$ to $\delta\tilde{g}=0.9$ in steps of $0.1$ from
  left to right). The $\delta$-peaks at $\langle   
J_z(\tau)\rangle/j=-1$ and 0 corresponding to $\delta\tilde{g}=0$ and
  $\delta\tilde{g}=1$ are not shown.}\label{fig.relax} 
\end{figure}

Figure~\ref{fig.relax} illustrates the incomplete relaxation in the inhomogeneous case for typical realizations of the stochastic process from Eq.~(\ref{SSE}).  For $\delta\tilde{g}=0$,
superradiance 
proceeds down to the ground state, whereas
for $\delta\tilde{g}=0.5$ the trajectories can get trapped at a random
finite value 
$\langle J_z(\tau\to\infty)\rangle$.  The statistical analysis of these
final values (we take $\tau=5$, as the trajectories typically have converged
at that time) leads to the histograms shown in the same figure, obtained
from $10^5$ runs of the SSE,  
$N_1=N_2=2$, and $N_1=N_2=3$. 
For $\delta\tilde{g}=0$, the histogram is a $\delta$-peak at $\langle
J_z(\tau)\rangle/j=-1$, as the superradiant relaxation proceeds 
to the total ground state.  For $\delta\tilde{g}=1$, a $\delta$-peak at $\langle
J_z(\tau)\rangle/j=0$ arises.  This is due to the fact that in this case the
second set of atoms has coupling constants zero. Therefore, these atoms
remain excited, whereas the atoms in the first set decay to the ground state,
such that in the end half of the excitation remains in the system, resulting
in $\langle
J_z\rangle=0$.  In general, for $N_1\ne N_2$ the final value for
$\delta\tilde{g}=1$ is given by
$\langle 
J_z(\tau\to\infty)\rangle/j=(N_2-N_1)/(N_2+N_1)$.   For intermediate values of
$\delta\tilde{g}$, the 
histogram 
rapidly broadens and shifts to larger values of  $\langle
J_z\rangle$ with increasing $\delta\tilde{g}$. 

\begin{figure}
\psfrag{xlabel}[b][b][1]{$\delta\tilde{g}$}
\psfrag{ylabel}[b][b][1]{$\overline{\langle J_z\rangle}/j$, $\sigma(\langle J_z\rangle)/j$}
\includegraphics[width=0.40\textwidth,clip=true]{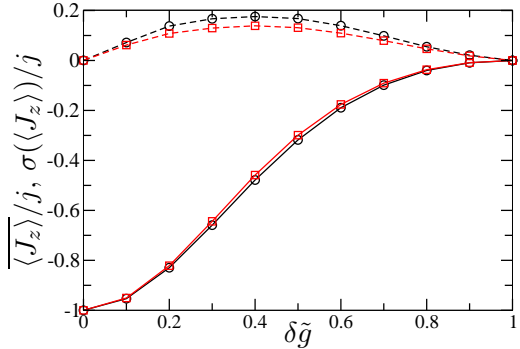}  
\caption{(Color online) Full lines: Average values 
  of $\overline{\langle J_z\rangle}/j$ 
  (over $10^5$ realizations of the SSE 
  Eq.~(\ref{SSE})) 
  and standard deviations (dashed lines) of $\langle
J_z(\tau)\rangle/j$ for large $\tau$ ($\tau=5$) as a function of
  $\delta\tilde{g}$. Black circles: 
  $N_1=N_2=2$; red/grey squares: $N_1=N_2=3$.  Full and dashed lines
  are guides to the eye.  
}\label{fig.relaxav}        
\end{figure}

Figure~\ref{fig.relaxav} shows the average value of  $\langle
J_z(\tau)\rangle/j$  for large $\tau$ ($\tau=5$) as a function of
$\delta\tilde{g}$ obtained from 
these $10^5$ realizations and the 
standard deviations (as errorbars) for $N=4$, $N_1=N_2=2$, and for
$N=6$, $N_1=N_2=3$. Initially we see quadratic behavior, which arises from
the vanishing  of $\langle
J_{z,1}(\tau)\rangle$ for  $N_1=N_2$ or $\tau\gg 1$.

\section{Conclusions}
We have presented a thorough analysis of the effect of superradiance of cold
atoms coupled to a superconducting on-chip $LC$-resonator.  Under realistic conditions we demonstrated a parameter regime in which
superradiance should be observable.  We have analysed the effect
of inhomogeneous couplings on the superradiance process by perturbation theory
in the inhomogeneity, and numerical simulations.  By dividing the sample into two subensembles, with different atom numbers and coupling constants, we can model the inhomogeneous coupling.  Our results show that
superradiance may be accelerated or slowed down compared to the homogeneous
case, depending  on the distribution of coupling constants.  The first order
correction in the inhomogeneity vanishes for all observables in the case of
two sub ensembles containing the same number of atoms.  For large times,
inhomogeneous coupling constants can lead to population trapping in random
decoherence free states, and we have provided numerical results for the
population inversion starting from a initially fully excited state.  The
first order correction to the final average population inversion vanishes in
this case, such that the change of that quantity is at most quadratic for small
inhomogeneity.

{\em Acknowledgments:} DB thanks the Joint Quantum Institute (University of
Maryland and NIST) for hospitality when this work was initiated, and CALMIP
(Toulouse) for the use of their computers.

\appendix
\section{Evaluation of matrix elements of angular momentum operators}
We use two different ways of calculating matrix elements of
angular momentum operators in the joint basis $|(j_1,j_2)jm\rangle$:
{\em i.)} Decoupling the basis states into single angular momentum basis states
using Clebsch-Gordan coefficients and {\em ii.)} using the
Wigner-Eckart theorem.
\subsection{Decoupling into single angular momentum basis states}
A straight forward way of obtaining the matrix elements of $J_{\pm}$
and $J_+J_-$ is to decouple 
\begin{equation} \label{u1}
\ket{(j_1,j_2)jm}=\sum_{m_1=-j_1}^{j_1}\CG{j_1}{m_1}{j_2}{m-m_1}{j}{m}\ket{j_1\,m_1}\ket{j_2\,m-m_1}\,,
\end{equation}
where the coefficients are Clebsch-Gordan coefficients. Then apply the
desired operator, and couple the states back together using 
the inverse transformation,
\begin{equation} \label{u2}
\ket{j_1\,m_1}\ket{j_2\,m-m_1}=\sum_{l=|j_1-j_2|}^{j_1+j_2}\CG{j_1}{m_1}{j_2}{m_2}{l}{m}\ket{(j_1,j_2)l\,m}
\end{equation}
Since the $J_\pm^{(i)}$ are symmetric under permutation of the atoms in
subsystem $i$, they conserve $j_i$, such that the matrix elements are all
diagonal in the index $\alpha=(j_1,j_2)$. We find
\begin{eqnarray}
\bra{(j_1,j_2)j\,m}\Jm{1}\ket{(j_1',j_2')j'\,m'}&=&\sum_{m_1=-j_1}^{j_1}d_-(j_1,m_1)\CG{j_1}{m_1}{j_2}{m+1-m_1}{j'}{m+1}\nonumber\\
&&\CG{j_1}{m_1-1}{j_2}{m+1-m_1}{j}{m}\delta_{j_1,j_1'}\delta_{j_2,j_2'}\delta_{m,m'-1}\label{j1mCG}\\
\bra{(j_1,j_2)j\,m}\Jp{1}\Jm{1}\ket{(j_1',j_2')j'\,m'}&=&\sum_{m_1=-j_1}^{j_1}d_+(j_1,m_1-1)d_-(j_1,m_1)\CG{j_1}{m_1}{j_2}{m-m_1}{j'}{m}\nonumber\\
&&\CG{j_1}{m_1}{j_2}{m-m_1}{j}{m}\delta_{j_1,j_1'}\delta_{j_2,j_2'}\delta_{m,m'} \label{j1pj1mCG}
\end{eqnarray}
Another derivation which in the end gives closed analytical expressions is
based on the Wigner-Eckart theorem.

\subsection{Wigner-Eckart theorem}
Let us consider momentarily a single angular momentum $j$ (i.e.~with
Hilbert space 
dimension $2j+1$ spanned by $\ket{jm}$ basis states which form a
simultaneous eigenbasis of $\bJ^2$ and $J_z$). The Wigner-Eckart
theorem states that the matrix elements  of an
irreducible  
tensor operator $T_{KQ}$ which transforms according to the irrep of $SU(2)$
with $j=K$, i.e.~like a state $\ket{KQ}$, is given by
\begin{equation} 
\bra{jm}T_{KQ}\ket{j'm'}=
(-1)^{2K}\bra{j}|T_K|\ket{j'}\CG{j'}{m'}{K}{Q}{j}{m}\,,  
\end{equation}
where $\bra{j}|T_K|\ket{j'}$ is a reduced matrix element that does not
depend on the magnetic quantum numbers $m$, $m'$ or $Q$ \cite{BrinkSatchler68}. In
practice one 
calculates these by using the Wigner-Eckart theorem backwards for a simple
operator $T_{KQ}$ whose matrix elements are known. There is just one scalar
tensor operator that can be formed from the components 
of  $\bJ$, $T_{00}(\bJ)\propto \bJ^2=j(j+1){\bf 1}$. Tensor operators of rank 1
(i.e.~a vector) are formed by the components of $\bJ$. We have
\begin{equation} \label{T1}
T_{1\pm 1}(\bJ)=\mp \frac{1}{\sqrt{2}}J_\pm,\,\,\,T_{10}(\bJ)=J_z\,.
\end{equation}
Higher order tensor operators of rank up to $K=k+k'$ can be formed from the
product of lower rank 
tensors $R_{kq}$, $S_{k'q'}$,
\begin{equation} \label{TRS}
T_{KQ}(R_k,S_{k'})=\sum_{q,q'}R_{kq}S_{k'q'}\CG{k}{q}{k'}{q'}{K}{Q}\,.
\end{equation}
One particular example is a tensor formed by the Cartesian product of the
components of the vector operator $T_{1q}(\bJ)$ ($q=-1,0,1$) introduced
above, which we denote as $T_{KQ}(\bJ,\bJ)$, and which reads
\begin{equation} \label{T2}
T_{KQ}(\bJ,\bJ)=\sum_{q,q'}T_{1q}(\bJ)T_{1q'}(\bJ)\CG{1}{q}{1}{q'}{K}{Q}\,.
\end{equation}
The Clebsch-Gordan coefficients limit the possible values of $K$ to
$K=0,1,2$. 
We can invert this relation and obtain the reduction of a product of
irreducible tensor operators into a sum of irreducible tensor operators,
\begin{equation} \label{prodT}
T_{1q}(\bJ)T_{1q'}(\bJ)=\sum_{l=0}^2\CG{1}{q}{1}{q'}{l}{q+q'}T_{l\,q+q'}(\bJ,\bJ)\,.
\end{equation}
With the help of this equation we can reduce the product $J_+J_-$ into its
irreducible components,
\begin{equation} \label{jpjm}
J_+J_-=-2T_{1+1}(\bJ)T_{1-1}(\bJ)=-2\left(\frac{1}{\sqrt{3}}T_{00}(\bJ,\bJ)
+\frac{1}{\sqrt{2}}T_{10}(\bJ,\bJ)+\frac{1}{\sqrt{6}}T_{20}(\bJ,\bJ)\right)\,. 
\end{equation}
Note that $T_{1,q}(\bJ,\bJ)=-T_{1,q}(\bJ)/\sqrt{2}$. This relation can be
shown component by component using Eqs.~(\ref{T2},\ref{T1}), and the
commutation relations of the angular momentum operators.\\

Now consider a composite system of two (physical or pseudo-) angular momenta
$j_1$ and $j_2$. We distinguish the operators acting on subsystem $i$ as
before by a superscript, $T_{KQ}^{(i)}$. The reduced matrix elements of an
operator which acts only  subsystem 1, $T_{KQ}=T_{KQ}^{(1)}\otimes {\bf
  1}^{(2)}$, can be related to the ones in subsystem 1 alone according to 
\begin{equation} \label{T1T}
\bra{(j_1,j_2)j}|T_K|\ket{(j_1',j_2')j'}=(-1)^{j+j_1'-K-j_2}\sqrt{(2j'+1)(2j_1+1)} W(j_1,j_1',j,j';K,j_2)\bra{j_1}|T_K^{(1)}|\ket{j_1'}\delta_{j_2,j_2'}\,,
\end{equation}
where the symbol $W$ is related to Wigner's $6j$-symbol by 
\begin{equation} \label{W}
W(a,b,c,d;e,f)=(-1)^{a+b+c+d}\left\{\begin{array}{ccc}
a&b&e\\
d&c&f
\end{array}
\right\}\,,
\end{equation}
see Eq.~(5.9) in \cite{BrinkSatchler68}. From (\ref{jpjm},\ref{T1T}) we obtain  
\begin{eqnarray}
\bra{(j_1,j_2)jm}\Jm{1}\ket{(j_1,j_2)j'm'}&=&\sqrt{2}(-1)^{j+j_1-j_2+1}\sqrt{(2j'+1)(2j_1+1)}\nonumber\\
&&W(j_1,j_1,j,j';1,j_2)
\CG{j'}{m'}{1}{-1}{j}{m}\bra{j_1}|T_1^{(1)}(\bJ)|\ket{j_1}\\   
\bra{(j_1,j_2)jm}\Jp{1}\Jm{1}\ket{(j_1,j_2)j'm'}&=&2(-1)^{j+j_1-j_2+1}\sqrt{(2j'+1)(2j_1+1)}\\
&&\Big\{\frac{1}{\sqrt{3}}
W(j_1,j_1,j,j';0,j_2)\CG{j'}{m'}{0}{0}{j}{m} 
\bra{j_1}|T_0^{(1)}(\bJ,\bJ)|\ket{j_1}\nonumber\\
&&-\frac{1}{\sqrt{2}}
W(j_1,j_1,j,j';1,j_2)\CG{j'}{m'}{1}{0}{j}{m} 
\bra{j_1}|T_1^{(1)}(\bJ,\bJ)|\ket{j_1}\nonumber\\
&&+\frac{1}{\sqrt{6}}
W(j_1,j_1,j,j';2,j_2)\CG{j'}{m'}{2}{0}{j}{m} 
\bra{j_1}|T_2^{(1)}(\bJ,\bJ)|\ket{j_1}\Big\}\,.\nonumber
\end{eqnarray}
It remains to calculate the reduced matrix elements. Brink and
Satchler \cite{BrinkSatchler68}
(p.51ff) give for a single angular momentum
$T_{00}(\bJ,\bJ)=\bJ^2/\sqrt{3}$. Using this in the Wigner 
Eckart theorem, sandwiched between states $\bra{j0}$ and $\ket{j'0}$ gives
for a single angular momentum
\begin{equation} \label{t0}
\bra{j}|T_0(\bJ,\bJ)|\ket{j}=-\frac{1}{\sqrt{3}}j(j+1)\,.
\end{equation}
Similarly, from $T_{10}(\bJ)=-J_z$ we find  
\begin{equation} \label{t11}
\bra{j}|T_1(\bJ)|\ket{j}=\sqrt{j(j+1)}\,,
\end{equation}
and 
\begin{equation} \label{t12}
\bra{j}|T_1(\bJ,\bJ)|\ket{j}=-\sqrt{j(j+1)/2}\,.
\end{equation}
Finally, one can show that $T_{20}(\bJ,\bJ)=(3J_z^2-\bJ^2)/\sqrt{6}$ by
applying 
(\ref{T2}), and the Wigner-Eckart theorem tells us that hence
\begin{equation} \label{t13}
\bra{j}|T_2(\bJ,\bJ)|\ket{j}=\sqrt{j(j+1)(2j-1)(2j+3)/6}\,,
\end{equation}
 and thus
\begin{eqnarray}
\bra{(j_1,j_2)jm}\Jm{1}\ket{(j_1,j_2)j'm'}&=&\sqrt{2}(-1)^{j+j_1-j_2+1}\sqrt{(2j'+1)(2j_1+1)j_1(j_1+1)}\nonumber\\
&&W(j_1,j_1,j,j';1,j_2)\CG{j'}{m'}{1}{-1}{j}{m}\label{Jp12}\\   
\bra{(j_1,j_2)jm}\Jp{1}\Jm{1}\ket{(j_1,j_2)j'm'}&=&2(-1)^{j+j_1-j_2+1}\sqrt{(2j'+1)(2j_1+1)}\label{JpJm12}\\
&&\Big\{-\frac{1}{3}
W(j_1,j_1,j,j';0,j_2)\CG{j'}{m'}{0}{0}{j}{m} j_1(j_1+1)\nonumber\\
&&+\frac{1}{2}W(j_1,j_1,j,j';1,j_2)\CG{j'}{m'}{1}{0}{j}{m} \sqrt{j_1(j_1+1)}\nonumber\\
&&+\frac{1}{6}
W(j_1,j_1,j,j';2,j_2)\CG{j'}{m'}{2}{0}{j}{m}
\sqrt{j_1(j_1+1)(2j_1-1)(2j_1+3)}\Big\}\,.\nonumber 
\end{eqnarray}
We have checked numerically up to $j=10$ and all combinations of
$j',j_1,m,m'$ (with $j_2=j'-j_1$) that Eqs.~(\ref{Jp12},\ref{JpJm12}) give
exactly the same results as Eqs.~(\ref{j1mCG},\ref{j1pj1mCG}), respectively.
Eqs.~(\ref{Jp12},\ref{JpJm12}) have the advantage of avoiding an additional
sum, and of leading to explicit expressions for the case relevant for the
calculation of $\langle J_z(\tau)\rangle$, as we demonstrate in Results section. Furthermore, the selection rule $j\in\{j'-2,\ldots,j'+2\}$ is evident
from these equations, as otherwise the Clebsch-Gordan coefficients
vanish.

\bibliography{../mybibs_bt}

\end{document}